%% file: main.tex
\documentclass{llncs}
\usepackage{cite}
\usepackage{amsmath,amssymb,amsfonts}
\usepackage{algorithmicx,algorithm,algpseudocode}
\usepackage{graphicx}
\usepackage{xcolor}
\usepackage{paralist}
\usepackage{float}
\usepackage[utf8]{inputenc}
\usepackage{authblk}
\usepackage{indentfirst}
\usepackage{paralist}
\usepackage{wrapfig}

\makeatletter
\renewcommand\section{\@startsection{section}{1}{\z@}%
                       {-8\p@ \@plus -4\p@ \@minus -4\p@}%
                       {6\p@ \@plus 4\p@ \@minus 4\p@}%
                       {\normalfont\large\bfseries\boldmath
                        \rightskip=\z@ \@plus 8em\pretolerance=10000 }}
\renewcommand\subsection{\@startsection{subsection}{2}{\z@}%
                       {-8\p@ \@plus -4\p@ \@minus -4\p@}%
                       {6\p@ \@plus 4\p@ \@minus 4\p@}%
                       {\normalfont\normalsize\bfseries\boldmath
                        \rightskip=\z@ \@plus 8em\pretolerance=10000 }}
\makeatother

\title{Cross Chain Bribery Contracts: Majority vs Mighty Minority
\thanks{An early version of the work was published in \textit{The PAAMS Collection: International Workshops of PAAMS 2019, Ávila, Spain, June 26 -- 28, 2019, Proceedings 17 (pp. 121-133).}}}
\author{Quang Tran \inst{1} \and Lin Chen \inst{2} \and Lei Xu \inst{3} \and Yang Lu \inst{1} \and Rabimba Karanjai \inst{1} \and Weidong Shi \inst{1}}
\institute{Computer Science Department, University of Houston, TX, USA \and 
Computer Science Department, Texas Tech University, TX, USA \and
Computer Science Department, Kent State University, OH, USA}

\begin{document}
\maketitle

\input{Abstract.tex}
\input{Intro.tex}

\input{Related_works.tex}
\input{Crosschain-incentives.tex}

\input{Case_Study.tex}

\input{Future_research.tex}

\input{Conclusion.tex}

\bibliographystyle{splncs04}
\bibliography{ref}
\end{document}

%% file: Abstract.tex
\begin{abstract}
Bribery is a perilous issue in the real world, especially in an economical aspect. This fraudulence is unavoidable, and more importantly, it is more difficult to trace in case smart contracts are utilized for bribing on a distributed public blockchain. In our paper, we propose a new threat to the security of a blockchain system, cross-chain bribery using smart contracts. An
arbitrary wealthy briber can utilize cross-chain smart contracts to manipulate a consensus mechanism on a victim’s blockchain or to disgrace a victim’s blockchain. To better understand this threat, our paper proposes a framework to analyze bribery using cross-chain smart contracts. We analyze the amount of incentive to bribe rational miners in a victim’s blockchain and also a full cost of conducting a cross-chain bribery attack. The result is that such attacks can be carried out with a reasonable amount of money or cryptocurrencies.
\end{abstract}

%% file: Intro.tex
\section{Introduction}
Blockchain \cite{nakamoto2008bitcoin, christidis2016blockchains, udokwu2018} provides a decentralized method for records keeping and information/transactions validation. 
Various applications are developed on top of the blockchain platform. One of these applications is a smart contract, which brings many advantages, e.g., saving time, reducing conflicts, and saving money. Several popular blockchain platforms support smart contracts such as Ethereum, EOS, Hyperledger Fabric, and Stellar, which can be feasible for solving a range of business challenges.~\cite{cointelligence_2018}. Even though smarts contracts provide a fundamental feature to extend usages of the blockchain, people tend to consider an actual usage of them while underestimating adverse effects they can cause. Smart contracts can be utilized in a destructive manner, i.e., bribery to undermine existing consensus mechanisms to gain financial benefits~\cite{DBLP:journals/corr/abs-1901-04620, DBLP:journals/corr/abs-1811-08263,inbook}. 

As one of the most popular blockchain construction method, the proof-of-work (PoW) based mining process requires a vast computational power to solve mathematical puzzles to produce a valid block. 
It is tough, if not impossible, for individual miners to compete with professional mining farm corporations. Therefore, individual miners usually join a mining pool to maximize their profits. In other words, we can claim that the miners in public blockchain are not united, and consolidated. They are forming up as a group based on the fundamental factor-maximizing efficiency and incentives. Thus, they can quickly change their mind and easily be manipulated, targeted in a bribery contract attack.

Unlike prior efforts primarily focusing on analyzing selfish mining behaviors using incentives restricted within a specific blockchain system to maximize a briber's beneficial rewards, we aim at outlining the possibility of security and stability of public blockchains and cryptocurrency systems as an economy driven game. We propose a cross-chain bribery attack scheme in which a briber targets at a distributed public blockchain to undermine consensus mechanism on a victim chain through a short-term bribing and manipulating bribed miners. The example scenario in our paper uses the Ethereum blockchain platform as a model to analyze how such an attack can be carried out. The cross-chain attack can be applied to any public blockchain since there always exists selfish behaviors within a public blockchain.

Our contributions in this paper include:
\begin{itemize}
	\item Providing a holistic view of selfish behaviors on blockchain by showing the feasibility of external influences in the form of cross chain bribery smart contracts.
    \item Investigating the effect of cross chain external incentives to consensus mechanism of victim blockchain.   
	\item Analyzing possible scenarios and consequences using selfish miners and validators in Ethereum blockchain as an example. 
	\item Providing a possibility of strategy by rational players who maximize profit in a multi-chain and multi-currency setting.
	\item Suggesting the necessity of additional research on selfish behaviors of rational miners in mining pools.
\end{itemize}

The rest of the paper is organized as follows. 
In section 2, we discuss selfish behaviors in blockchain and related works. Then we introduce cross chain incentives in the form of bribery smart contracts and analyze their influences on selfish behaviors on the victim chain. After the analysis, we discuss future research direction and open problems. In the end, we conclude the paper.

%% file: Related_works.tex
\section{Selfish Mining Strategy and Related Works}\label{sec-related}
\vspace{-15pt}
\begin{figure}[H]
  \centering
  \includegraphics[width=7.5cm]{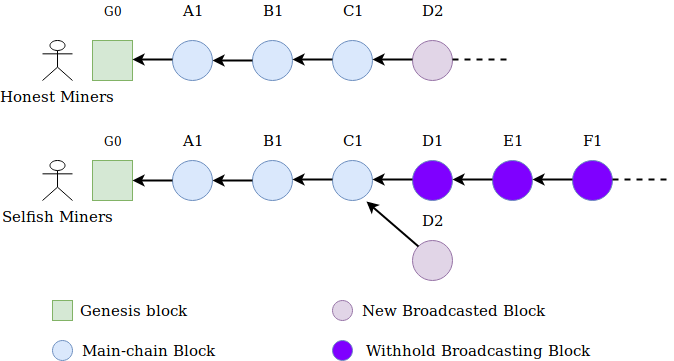}
  \caption{Selfish miners withhold broadcasting new blocks.}
  \label{fig-Selfish}
\end{figure}
\vspace{-15pt}
In a blockchain system built on proof-of-work (PoW) and longest chain principle, a regular user works on the latest block on the longest branch to produce a new block. After the user successfully makes a valid block, he/she broadcasts the new block to the whole network immediately to claim the reward.
However, selfish miners who want to maximize their benefits can adopt a different strategy to get more rewards. One possible strategy is to hold produced blocks and submits them together to the system.

The idea of selfish mining gains more attention since it was first proposed in 2014~\cite{EyalSirer2014}. The paper pointed out that a simple majority is not enough to protect the security of Bitcoin in a case of the existence of selfish miner. Specifically, they propose an attack that selfish miners can push their revenue up while gaining more rational miners to join the selfish mining pool until it becomes a majority. 
Following this inspiration, other models are proposed to stir up the attention of researchers and developers in understanding such a potential threat of selfish mining in a public blockchain platform ~\cite{DBLP:journals/corr/abs-1901-04620, DBLP:journals/corr/abs-1811-08263}. Jian Niu and Chen Feng analyzed selfish mining in the Ethereum platform using a 2-dimensional Markov model to determine a threshold that makes such fraud strategy profitable, which is lower than the one in Bitcoin. However, in reality, the chance of selfish mining occurrence can be meager, or it might not happen. Two essential conditions need to be met which are computational power, and hashing power. Selfish miners must have a strong computational power and control enough hashing power so that they can generate blocks quicker than honest and majority miners. It becomes a game of cat and mouse.

Besides controlling computation to perform selfish mining, a new methodology, known as a bribery contract, is also considered which can achieve similar effects~\cite{Kothapalli2016ABF,mccorry_hicks_meiklejohn_2019 }. A smart contract is an application written in a programming language by developers running on top of a supported blockchain platform. A smart contract contains a set of rules proposed by a creator to those who interact and accept these rules. When these pre-defined conditions are met, a contract agreement is automatically enforced, and a transaction is also automatically generated to a network to verify before inserting it into a blockchain. A smart contract feature can cause an impact on the current incentive and fairness mechanisms in a blockchain consensus protocol. Selfish miners make use of electronic contracts to create a bribery attack on a targeted blockchain platform by giving bribees incentive rewards for doing fraudulence. One of an example is that briber creates a bribery contract to give an amount of coin token as a reward for those miners who withhold announcing new successful mining block until briber successfully mines and announces this block to a network. This attack can be achieved easily by any wealthy adversary without the need of competing hashing and computational power with existing miners. Importantly, this fraudulence favors both the briber and bribee since they both get some profit for doing so, especially in the case of the Ethereum platform. A briber gets a block mining reward while a bribee also gets both uncle block reward and an additional bribery contract reward.

%% file: Crosschain-incentives.tex
\section{Selfish Behavior with Cross-chain Incentives}\label{sec-crosschain-bribery}
\input{sub-bribery-contract.tex}

\input{sub-influence-outside.tex}

%% file: sub-bribery-contract.tex
\subsection{Bribery Contracts - the New Threat to Consensus}
Blockchain has been used to build various applications across many business platforms \cite{8038514,xu_chen_shah_gao_lu_shi_2017}, and the smart contract is a fundamental feature to extend blockchain's impacts~\cite{10.1007/978-3-319-70278-0_29,XuShahChenNourZhimin2017,karamitsos_papadaki_barghuthi_2018}. 
Most existing works consider an actual usage of smart contracts, but tend not to pay attention to the other side of it, i.e., smart contracts can be used in a destructive manner. 
In particular, smart contracts can be utilized for bribery to affect existing consensus mechanisms. 
Bribery attacks or bribery contracts are severe problems like the fact that it can manipulate or destroy the fundamental assumption of standard smart contract execution model which primarily relies on consensus or majority accepted outcome~\cite{Chen2017SmartCE,inbook}. An arbitrary person participates in a game and accepts game's rules does not necessarily mean that this person will never be manipulated to change his/her mind when he/she is offered an appealing compensation for violating game's rules. This attack can be achieved easily by any wealthy adversary. 
Remarkably, tracing a briber in a blockchain system seems to be extremely hard since it is designed as a decentralized environment and to protect user anonymity \cite{heilman_baldimtsi_goldberg_2016}. Importantly, the fraudulence behavior favors both briber and bribees since they both achieve some goals after.

%% file: sub-influence-outside.tex
\subsection{Influences from Outside of a Blockchain}
\vspace{-15pt}
\begin{figure}[H] 
    \centering
        \includegraphics[width=0.5\textwidth]{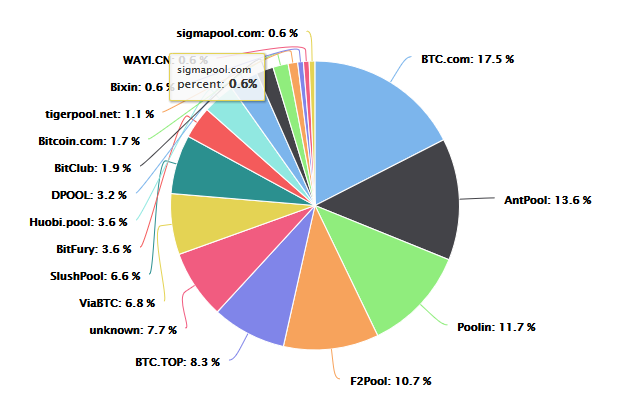}%
        \label{fig:a}%
    \hfill%
        \includegraphics[width=0.5\textwidth]{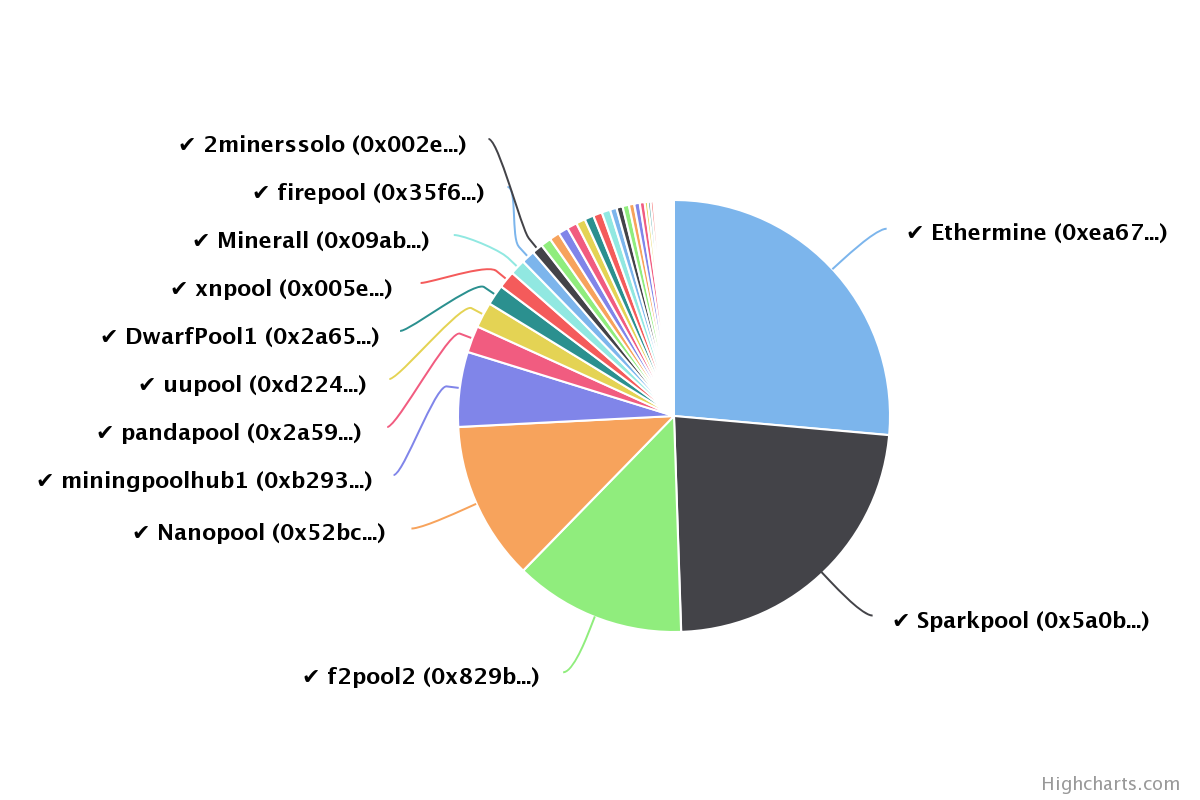}%
        \label{fig:b}%
    \caption{Bitcoin and Ethereum mining pool hash rate power}{Captured from \footnotemark}
\end{figure}
\vspace{-15pt}
\footnotetext{https://btc.com/stats/pool} 
\footnotetext{https://www.etherchain.org/charts/topMiners}
A distributed blockchain is built on the platform of many algorithms, security protocols which aim at maintaining unity and fairness for all joining parties. While the algorithms, protocols, and other factors in the blockchain network are fixed and are not likely to change, the human behavior of users in the system is complicated and may be prone to change. Unlike a machine which operates exactly as it is programmed, many people may change their behavior based on the surrounding conditions (and thus deviate from the protocol), and this is mainly the case for those who are chasing economic benefits in a system of financial incentives like a blockchain. 

The Fig. 2 shows the power hash rate of Bitcoin and Ethereum mining pools. Note that these mining pools are not operated by individual ones. Instead, these are groups of miners who join a mining pool to get a better profit. In other words, these can be rational miners who intentionally chase the profit in mining. If they are offered a better amount, there is no doubt that they can easily switch.

"Why buy when you can rent?"~\cite{10.1007/978-3-662-53357-4_2} since there always exist rational miners in a distributed-public blockchain. Note that our paper is aimed to discuss on possibility of cross-chain bribery attack. A briber or a group of briber targets and attacks a public blockchain on purpose. They are whales and have enough fund to perform an attack on another blockchain in favor of controlling block generation or disgracing another blockchain. Remarkably, the briber does not require to have a majority of hashing power nor to participate in mining. They can create a cross-chain contract to fascinate those rational miners in a targeted blockchain network.

%% file: Case_Study.tex
\section{Case Study and Analysis}\label{sec-examples}
In this section, we provide an example scenario of cross-chain bribery contract using decentralized-base platform Ethereum as a model architecture and victim chain. Note that the scenario can be applied on any public blockchain platform.
\input{sub-introduce-Ethereum}
\input{sub-structure}
\input{sub-reward}
\input{sub-pseudocode}
\input{sub-incentive}

%% file: sub-introduce-Ethereum.tex
\subsection{Ethereum Platform}
Ethereum \cite{ethereum}, a second largest decentralized cryptocurrency platform by market capitalization \footnote{Available from:  https://coinmarketcap.com/all/views/all/}, allows for the execution of smart contracts on the blockchain. To create a protocol for smart contracts which offer beneficial and efficient interactions between participants in the network, Ethereum builds a Turing-complete machine, called as Ethereum Virtual Machine or EVM, as the heart of its blockchain platform. Developers can create their application, run on EVM using any friendly programming languages, to create their own arbitrary rules of ownership, transaction formats, and state transition functions. Thus, in term of a smart contract, Ethereum can sometimes be considered as a "world computer". 

In Ethereum, a user is called a client. A client that runs mining on Ethereum blockchain is called a miner/a node. A client can send ETH, which is the cryptocurrency of Ethereum, to a smart contract or to other clients by issuing a transaction. Every transaction that is deployed costs some gas fee to execute. The gas fee is an incentive reward to a miner who collects those transactions and attaches them into a new mining block. On Ethereum blockchain, a new mining block can be an empty block or a block that contains a number of transactions which are limited by "gas limit". Moreover, there is no centralized party to validate new mining blocks. By default, one node can connect up to 25 peers in the network to form up a subset of nodes. Many subsets of nodes in the Ethereum network take responsibility for broadcasting and validating new mined blocks when they are announced.

%% file: sub-structure.tex
\subsection{Ethereum Blockchain Structure}
In Ethereum, there are two types of blocks - an uncle block and a main block. A main block is a valid block and is appended to a longest chain. Unlike Bitcoin platform which does not accept a late broadcasting block as an uncle block, Ethreum offers this feature to maintain the security of the chain, which allows for faster block generation time ($\approx$15 seconds in Ethereum, and $\approx$10 minutes in Bitcoin). A standard Ethereum's block structure consists of three components: a block header (contains parent's block hash, account's address, Merkle Tree root hash, a time stamp, block's difficulty, and a nonce,...), list of references to uncle blocks (max uncle reference is 2), and a set of transactions. While a main block contains three components as above, an uncle block only contains a block header. To be considered as an uncle block, this block requires to be referenced by another next main block within 7 rounds (see \figurename~\ref{fig-Calculate-Unclue-Reward}). Otherwise, it will be considered as "block lost".

%% file: sub-reward.tex
\subsection{Mining Reward}
In Bitcoin blockchain, because block generation time is high ($\approx$10 minutes) which overcomes block propagation delay, an orphan block-a block is broadcast after a main block- is discarded and is not given any reward.
Unlike Bitcoin, Ethereum aims to increase transaction's throughput by decreasing a block generation time ($\approx$15 seconds). Thus, to maintain the security of its chain, a late broadcasting block is also accepted as an uncle block and is given a partial reward. Hence, there are three types of block rewards: a main block reward, an uncle block reward, and a nephew block reward \cite{ethereumMining, Nakamoto2015ANG}. The main block reward is being used in both Bitcoin and Ethereum, which gives a reward to encourage those miners to solve a computational puzzle as fast as possible. The uncle and nephew block rewards are exclusive rewards by Ethereum. Because uncle blocks are submitted later than the main block, they are only given a partial reward which depends on when another main block references them. And the main block which references the uncle block is also given an additional reward, which is a nephew reward, to encourage miners attaching uncle blocks to maintain the chain's security.

\begin{table}[H]
\resizebox{\textwidth}{!}{%
\begin{tabular}{lccl}
\hline
                    & \qquad Ethereum \qquad & \multicolumn{1}{l}{Bitcoin \qquad} & Usage                                                                        \\ \hline
Main Reward         & Yes      & Yes                         & Incentive mining block                                                       \\ \hline
Uncle Reward        & Yes      & No                          & Maintain chain's security                                                    \\ \hline
Nephew Reward       & Yes      & No                          & Encourage to reference uncle block                                           \\ \hline
Bribe Uncle Reward  & No       & No                          & Special reward in our theory giving to miner who mines on fork chain         \\ \hline
Bribe Accepted Reward & No       & No                          & Special reward in our theory giving to nodes who accept blocks on fork chain \\ \hline
\end{tabular}%
}
\vspace{2pt}
\caption{Mining Reward}
\label{my-label}
\end{table}
\vspace{-15pt}
Each miner who successfully mines a main block can receive a reward of 2.0 ETH (not include nephew rewards, transactions fee reward). The uncle block reward is not a fixed number. The amount of reward is various since an uncle block is required to be referenced by a later main block. The sooner it is referenced, the higher the reward it is given. An equation to calculate an uncle block reward as:
\centerline{$U_R = (U_n + 8 - B_n) * R / 8$} 
\begin{flushleft}
$U_R$: uncle reward \qquad $U_n$: Uncle block height \qquad $R = 2.0$ ETH \\
$B_n$: referenced by main block's height 
\end{flushleft}
\vspace{-15pt}
\begin{figure}[H]
  \centering
  \includegraphics[width=12cm, height=4cm]{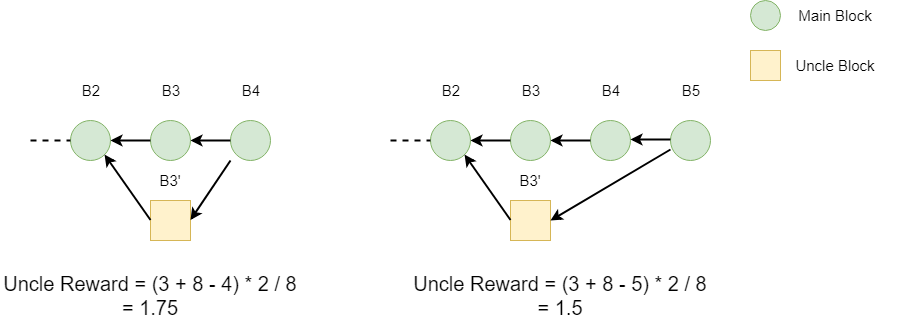}
  \caption{Uncle block reward varies by time}
  \label{fig-Calculate-Unclue-Reward}
\end{figure}
\vspace{-15pt}

In our theory scheme, we propose two more additional rewards to achieve a bribery attack: a bribe uncle reward and a bribe acceptance reward. The bribe uncle reward is given to those miners who accept to mine a block on a fork-chain instead of a longest chain. The bribe accepted reward is a reward giving to those clients/nodes which accept and insert a block on a fork-chain into their blockchain. The amount of each reward is discussed later in the following section.

%% file: sub-pseudocode.tex
\subsection{Bribery Contract and Pseudocode}
In our scheme of cross-chain bribery attack, we use the Ethereum blockchain platform as a model architecture to provide an example scenario. Note that this scheme can be applied on any public blockchain which is targeted by the briber. To set up our scheme, we consider the case of one briber who targets Ethereum to perform the bribery cross-chain attack. The briber has an account with his/her crypto-coins on another blockchain platform and also has enough money to perform this attack. Remarkably, a briber can be any individual-wealthy person or a group of people. They do not necessarily have a majority or even own any of Ethereum's computational power to do this bribery cross-chain attack. Their goal is to manipulate miners/nodes in the Ethereum network to control the generation of blocks or to disgrace Ethereum blockchain.

Miners run the mining process to find a block in getting an incentive reward. They can be an individual miner or can join a mining pool to pursue a more beneficial reward. Notably, they all share a common concept - maximizing the efficiency and profit of their mining process. More importantly, even if a miner is honest in the classical case, it does not necessarily mean they remain to be honest when they are offered a better compensation \cite{inbook, DBLP:journals/corr/abs-1901-04620, DBLP:journals/corr/abs-1811-08263, EyalSirer2014}. Thus, a bribery cross-chain contract is feasible when both a briber and bribees can achieve their goals. 

In our scenario, we propose a $BribeContract$ that rewards miners who intentionally mine blocks on a fork-chain, and also rewards bribees who accept and insert this new mining block into their chain. These pre-defined rules are created by the briber. Furthermore, a transaction reward is automatically issued when bribees prove that they meet the requirements. Our proposed approach requires one briber to pay a full cost to perform this cross-chain attack. The amount of the bribe incentive reward should be enough appealing so that bribees are willing to join. We discuss more on this in the following section.\\

\begin{enumerate}
    \item \textbf{Briber creates a BribeContract on another blockchain platform.} The BribeContract specifies the rules for the attack, such as the amount of the bribe incentive reward, the criteria for miners to be considered as bribees, and the process for issuing the transaction reward.
    \item \textbf{Briber sends the bribe incentive reward to the BribeContract}. The bribe incentive reward is locked in the BribeContract and cannot be withdrawn by the briber.
    \item \textbf{Miners mine blocks on a fork-chain of the target blockchain.} The miners are incentivized to mine blocks on the fork-chain because they will receive the bribe incentive reward if their blocks are accepted by the bribees.
    \item \textbf{Bribees accept and insert the new mining blocks into their chain.} The bribees are incentivized to accept and insert the new mining blocks into their chain because they will receive the transaction reward.
    \item \textbf{The target blockchain is forked.} The new mining blocks on the fork-chain are accepted by the bribees and become part of their chain. This creates a fork in the target blockchain.
    \item \textbf{The briber achieves their goal.} The briber's goal may be to manipulate the miners/nodes in the target network to control the generation of blocks or to disgrace the target blockchain.
\end{enumerate}

\noindent \textbf{Pseudocode:}
\setlength{\intextsep}{5pt}
\begin{algorithm}[H]
\caption{Propose Contract}
\begin{algorithmic}[1]
\Procedure{Bribe Contract}{} \\
- Briber $B$ creates a bribe contract $\beta$ \\
- Contract $\beta$ contains: \\
\indent + $\alpha_{a}$: contract creator's address \\
\indent + $\mu_m$: amount incentive for mining block on fork-chain \\
\indent + $\mu_a$: amount incentive for accepting mining block on fork-chain \\
\indent + $\gamma$: total side-chain token deposits into the contract \\
\indent + $f(b)$: a context to identify arbitrary bribee (b) \\
\indent + $f(c)$: a context function of condition (c) is met \\
\indent + $f(t)$: a context to terminate contract when a fork-chain becomes a main chain. \\
\indent + $S_b$: starting block on fork chain.
\EndProcedure
\end{algorithmic}
\end{algorithm}

\begin{algorithm}[H]
\caption{Prove and Commit}
\begin{algorithmic}[2]
\Procedure{Prove}{} \\
- Bribee (b) proves contract's condition (c) is met
\If {$c =$ \textit{mining block on fork chain}}
    \State $b$ commits a fork-chain to contract $\beta$
\EndIf
\If {$c =$  \textit{accepting block on fork chain}}
    \State $b$ commits a fork-chain to contract $\beta$
\EndIf
\EndProcedure
\end{algorithmic}
\end{algorithm}
\setlength{\intextsep}{5pt}
\begin{algorithm}[H]
\caption{Verify}
\begin{algorithmic}[3]
\Procedure{Verify}{} \\
- A block should stay within a range of required number blocks \\
- For example: $S_b$ ... $\leftarrow$ $U_{v}$ $\leftarrow$ $U_{v+1}$ $\leftarrow$ $U_{v+2}$...$\leftarrow$ $U_{v+7}$
\If {$c =$ \textit{mining block on fork-chain}}
    \If {$U_v \in [S_b, U_{v+7}]$}
        \State $b \gets \mu_m$
    \EndIf    
\EndIf
\If {$c =$  \textit{accepting block on fork-chain}}
    \If {$U_v \in [S_b, U_{v+7}]$}
        \State $b \gets \mu_a$
    \EndIf    
\EndIf
\EndProcedure
\end{algorithmic}
\end{algorithm}
\setlength{\intextsep}{5pt}
Algorithm 1 presents the contract proposal. This contract includes all required variables, functions to detect, to verify and to give incentive reward to bribees when a contract's condition is qualified. Algorithm 2 shows the process of a bribee, proving how he/she is qualified to receive a reward. If a bribee mined a block on a fork-chain, he/she should send its chain which is started from block $S_b$ up to the latest block on a fork-chain. The similar request is also required in case of the bribees accepting and inserting a mined block on a fork-chain. Due to that Ethereum's chain structure design is tamper-free, a verifying block to be submitted must stay within a range of multiple blocks on the fork-chain. Algorithm 3 verifies this requirement to give a corresponding reward to bribees.

%% file: sub-incentive.tex
\subsection{Incentive Discussion}
The cross-chain bribery framework we described before needs to encourage both bribers and bribees to participate in such an attack. The incentive mechanism consists of two parts:
\begin{itemize}
    \item Incentive reward that bribees receive when participating in a cross-chain bribery attack.
    \item The cost that a briber must pay so that they can achieve his/her goals.
\end{itemize}

\noindent\textbf{Bribees}: Although miners can be honest in the traditional mining process (without bribery), it does not necessarily mean they remain to be honest when they are offered a better compensation. The amount of Ethereum token that one miner receives can be described as:

$$\theta_h = \frac{\alpha}{\beta} \times R \times \frac{3600}{\phi},$$
where:
\begin{inparaenum}[(i)]
    \item $\theta_h$ denotes a number of receiving tokens in one hour
    \item $\alpha$ denotes a hashing power of hardware using to solve a computational puzzle
    \item $\beta$ denotes a network hash rate
    \item $R$ denotes a main block reward ($R = 2$ ETH)
    \item $\phi$ denotes a block generation time
\end{inparaenum}
~\cite{ethereum_network_status}.

Unlike Bitcoin, it is less common to use ASIC to mine Ethereum. 
Most miners use GPU or CPU for mining. 
Thus, we set the value $\alpha$ in the range $[10, 400]$ MHash/s in our model.

\begin{figure}[H] 
    \centering
        \includegraphics[width=0.5\textwidth]{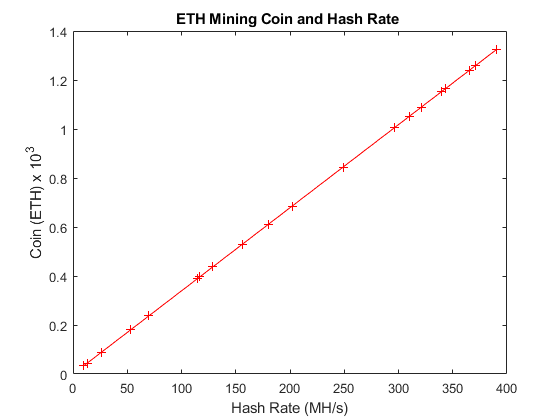}%
        \label{fig:hourly:reward:a}%
    \hfill%
        \includegraphics[width=0.5\textwidth]{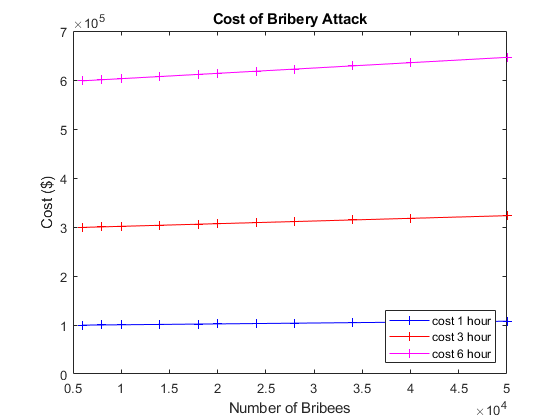}%
        \label{fig:hourly:reward:b}%
    \caption{Hourly reward and bribery contract cost}
    \label{fig:hourly:reward}
\end{figure}

\begin{table}[H]
\centering
\resizebox{0.8\textwidth}{!}{%
\tiny
\begin{tabular}{|l|c|c|c|}
\hline
Number of Bribees & \multicolumn{1}{l|}{1 Hour (\$)} & \multicolumn{1}{l|}{3 Hours (\$)} & \multicolumn{1}{l|}{6 Hours (\$)} \\ \hline
10,000            & 100,400                          & 301,300                           & 602,500                           \\ \hline
20,000            & 102,200                          & 306,600                           & 613,300                           \\ \hline
30,000            & 104,000                          & 312,000                           & 624,000                           \\ \hline
40,000            & 105,800                          & 317,300                           & 634,700                           \\ \hline
50,000            & 108,000                          & 324,000                           & 648,000                           \\ \hline
\end{tabular}%
}
\vspace{5pt}
\caption{Cost of cross-chain bribery attack}
\label{tbl:cost:attack}
\end{table}

\figurename~\ref{fig:hourly:reward} (left) shows the hourly reward that honest miners can receive by their contribution of a hash rate in mining a new block. The maximum ETH token that miners ($\alpha = 400$ MHash/s) can receive in one hour is around $\theta_h = 0.0014$ ETH/h. If a briber offers these rational miners a better incentive amount (i.e. $\tau_h = 0.002$ Token-worth(\$) in cross-chain for accepting new blocks, and $\gamma = 3.0$ Token-worth(\$) for mining on the fork-chain), rational miners have a high possibility to join our bribery cross-chain attack. The incentive bribe reward that bribees-rational miners in mining pool-can get if they join the attack as:
$$\Gamma = \tau_h + \frac{\alpha}{\beta} \times 3 \times \frac{3600}{\phi}.$$

\noindent \textbf{Briber}: A briber can be any wealthy-individual person or a group of people. They have enough fund to perform a cross-chain bribery attack. Their goal is to control blocks' generation or to disgrace Ethereum blockchain by bribing rational miners/nodes in the network. The major issue is how much it costs to perform such an attack.  \figurename~\ref{fig:hourly:reward} (right) and \tablename~\ref{tbl:cost:attack} show that the budget of cross-chain bribery attack is acceptable to bribers. \\

\noindent \textbf{Consequences:} In our model, a block generation time $\phi$ is around 15 seconds. Within six hours, the number of blocks can be generated by bribed miners is around 1440 blocks. If a briber targets one public blockchain to deploy cross-chain bribery attacks within a short time (i.e. 6 hours of attack), their cost (see \tablename~\ref{tbl:cost:attack} and \figurename~\ref{fig:hourly:reward}) to perform such an attack is lower than the potential damage caused to a victim blockchain. The attacker can potentially undermine a consensus mechanism of a victim chain or secretly shot a price of the victim chain currency while gaining financial benefits of dropped price or aim to disgrace the targeted chain.

%% file: Future_research.tex
\section{Future Research and Open Problems}

Our work points to a new direction of analyzing the security and stability of public blockchains and cryptocurrency systems. Distinguishing from prior efforts primarily focusing on analyzing selfish behaviors using incentives confined within a blockchain system (e.g., block reward), our approach applies a holistic view of of systems where a network of public blockchains and crypto-currencies are considered as inter-connected systems where external influences can happen in the form of cross-chain transactions and contracts. This may significantly change the selfish behaviors of users, and affect our understanding of sustainability and stability of public blockchain systems. 

Our preliminary results suggest the necessity of additional research, in particular, detailed modeling of decisions and strategies that may be adopted by rational players in the existence of cross-chain incentives for selfish behaviors, risk posed by cross-chain bribery contracts to stability and well-being of victim chains, holistic analysis of strategy by rational players who maximize profit in multi-chain and multi-currency setting, and possible mitigation strategies and design options. 

Holistic analysis under the agent perspective raises many open problems with respect to the nature of players in public blockchains. Although researchers are amenable to introduce rational players in analyzing public blockchains, there is a lack of consensus on proper assumptions of the players and implications of the assumptions to long term sustainability of public blockchain systems. In the presence of rational players with in-chain behaviors influenced by cross-chain rewards, how does it affect a theoretical analysis of blockchain consensus and security properties?  In particular, an adversary can potentially gain profit from one chain by deliberately inducing disruptions to other chains. It means that selfish players who engage in bad behaviors within a blockchain may not necessarily look for rewards within the same blockchain.

%% file: Conclusion.tex
\section{Conclusion}\label{sec-conclusion}
Blockchain has found a variety of applications and it is critical to guarantee the correctness of the system. We study a new threat to the security of a blockchain, cross-chain bribery using smart contracts. Bribery is a perilous problem in the real world, especially in an economical aspect. It is unavoidable fraud and more importantly, difficult to find since it is utilized by cross-chain smart contracts on the distributed public blockchain. Recent studies have shown corrupted fraud utilizing smart contracts to conduct bribery. In our paper, we improve this idea by proposing a cross-chain bribery attack to undermine a victim's consensus mechanism. In this paper, we outline the possibility of a cross-chain bribery attack by bribing selfish, rational miners in a targeted network and discuss the potential example scenario. A cross-chain bribery attack is feasible to facilitate on public blockchains and cryptocurrency systems due to the fact that there always exist rational miners who are incentivized by beneficial rewards. The possibility of all miners avoid short term benefit to protect a long term one seems to be negligible. People might realize they did harmful things to others. However, in terms of getting a better beneficial reward, no one would even doubt trying to do such a thing. Remarkably, the cost of carrying out one cross-chain bribery attack can be acceptable while it can cause tremendous on a victim's chain such as dropping the price, or disgracing victim's blockchain.